\RequirePackage{fix-cm}
\documentclass[smallextended]{svjour3}       
\usepackage[utf8]{inputenc}
\usepackage{xspace}

\smartqed  
\usepackage{graphicx}
\journalname{Journal of Statistical Physics}

\usepackage{amsmath}
\usepackage{amssymb}
\usepackage{mathtools}
\usepackage{hyperref}
\usepackage[caption=false]{subfig}

\newcommand\subfig[2]{{Fig.~\ref{#1}{#2}}}
\newcommand\subcap[1]{{(#1):}}

\newcommand{\SET}[1]{\{#1\}}

\newcommand{\sub}[1]{\texttt{#1}}
\newcommand{\eq}[1]{eq.~\eqref{#1}}
\newcommand{\eqtwo}[2]{eqs~\eqref{#1} and~\eqref{#2}}

\newcommand{\fig}[1]{Fig.~\ref{#1}}
\newcommand{\figs}[2]{Figs~\ref{#1} and~\ref{#2}}
\newcommand{\quot}[1]{``#1''}
\newcommand{\tab}[1]{Table~\ref{#1}} 
 
\newcommand{\sect}[1]{Section~\ref{#1}} 
\newcommand{\subsect}[1]{Subsection~\ref{#1}} 
\newcommand{\subsecttwo}[2]{Subsections~\ref{#1} and~\ref{#2}} 
\newcommand{\app}[1]{Appendix~\ref{#1}}

\newcommand{\etc}{\textrm{etc.}}
\newcommand{\etcp}{\textrm{etc}}
\newcommand{\ie}{\textrm{i.e.}}

\newcommand{\ACAL}{\mathcal{A}}  
\newcommand{\BCAL}{\mathcal{B}}  
\newcommand{\MCAL}{\mathcal{M}}  
\newcommand{\OCAL}{\mathcal{O}}  
\newcommand{\eps}{\varepsilon}
\newcommand{\expb}[1]{\exp \glb #1 \grb} 
\newcommand{\ranb}[2][]{\sub{ran}_{#1} \! \glb #2 \grb}  

\newcommand{\loga}[2][]{\log^{#1}\! \gla #2 \gra}  

\newcommand{\minc}[2][]{\min^{#1} \glc #2 \grc}  

\newcommand{\gla}{\,}  
\newcommand{\gra}{}  
\newcommand{\glb}{\left(}  
\newcommand{\grb}{\right)}  
\newcommand{\glc}{\left[}  
\newcommand{\grc}{\right]}  

\newcommand{\const}{\text{const}}

\newcommand{\TO}{,\ldots,}
\newcommand{\VEC}[1]{\mathbf{#1}}
\newcommand{\xvec}{\VEC{x}}

\newcommand{\alphavec}{\boldsymbol{\alpha}}

\newcommand{\deltavec}{\boldsymbol{\delta}}
\newcommand{\Deltavec}{\boldsymbol{\Delta}}

\newcommand{\epsvec}{\boldsymbol{\varepsilon}}

\newcommand\bigOb[1]{\ensuremath{\OCAL\glb #1 \grb}}

\newcommand{\fpn}[2]{\ensuremath{#1 \! \times \! 10^{#2}}}

\newcommand{\TIME}{t}
\newcommand{\ESCAPE}{\TIME_\mathrm{esc}}
\newcommand{\MAXNEAR}{d}
\newcommand{\MAXNEARBOUND}{\gamma}
\newcommand{\BP}{Böröczky packing\xspace}
\newcommand{\BPS}{Böröczky packings\xspace}

\newcommand{\EBC}{$\eps$-relaxed Böröczky configuration\xspace}
\newcommand{\BCORE}{Böröczky core\xspace}
\newcommand{\KCORE}{Kahle core\xspace}

\newcommand{\EBCS}{$\eps$-relaxed Böröczky configurations\xspace}
\newcommand{\LSP}{locally stable packing\xspace}

\newcommand{\DIST}[2]{\text{dist}(#1,#2)}
\newcommand{\tchain}{\tau_{\text{chain}}}
\newcommand{\tcpu}{t_{\text{cpu}}}
\newcommand{\TMCMC}{t_{\text{MCMC}}}
\newcommand{\DENSITY}[3]{$#1\ \pm \ \fpn{#2}{#3}$}
\newcommand{\ACALGEO}{\ACAL^{\text{geo}}}
\newcommand{\ACALCIRC}{\ACAL^{\text{circ}}}
\newcommand{\attenuation}{\phi}
\newcommand{\Vcut}{V_{\text{cut}}}
\newcommand{\Lcut}{L_{\text{cut}}}
\newcommand{\MESCAPE}{\MCAL^{\text{esc}}}
\newcommand{\Veq}{V^{\text{eq}}}

\begin{document}

\title{
Sparse hard-disk packings and local Markov chains \thanks{P.H. acknowledges 
support from the Studienstiftung des deutschen Volkes and from Institut 
Philippe Meyer. W.K. 
acknowledges support from the Alexander von Humboldt Foundation.}  
}


\author{Philipp H\"ollmer
   \and Nicolas Noirault \and Botao Li \and A. C. Maggs \and Werner Krauth
}


\institute{
Philipp H\"ollmer \\ 
Bethe Center for Theoretical Physics, University of Bonn, Germany \\
\email{hoellmer@physik.uni-bonn.de}\\
\and
Nicolas Noirault\thanks{abc}\\ 
Laboratoire de Physique de l’Ecole normale supérieure, ENS, 
Université PSL, CNRS, Sorbonne Université, Université de Paris,
Paris, France\\
\email{nicolas.noirault@laposte.net}\\
\and
Botao Li \\
Laboratoire de Physique de l’Ecole normale supérieure, ENS, 
Université PSL, CNRS, Sorbonne Université, Université de Paris,
Paris, France\\
\email{botao.li@phys.ens.fr}\\
\and 
A. C. Maggs\\
CNRS Gulliver, ESPCI Paris, Université PSL, 10 rue Vauquelin, 75005 
Paris, France. \\
\email{anthony.maggs@espci.fr}\\
\and
Werner Krauth \\
Laboratoire de Physique de l’Ecole normale supérieure, ENS, 
Université PSL, CNRS, Sorbonne Université, Université de Paris,
Paris, France\\
\email{werner.krauth@ens.fr}\\
}

\date{Received: date / Accepted: date}

\maketitle
\setcounter{tocdepth}{3}

\begin{abstract}
We propose locally stable sparse hard-disk packings, as introduced by Böröczky, 
as a 
model for the analysis and benchmarking of Markov-chain 
Monte Carlo  (MCMC)
algorithms. We first generate such packings in a square box with periodic 
boundary conditions and analyze their properties. We then study how local 
MCMC algorithms, namely the Metropolis algorithm and several 
versions 
of 
event-chain Monte Carlo (ECMC), escape from configurations that are obtained by 
slightly reducing all disk radii by a relaxation parameter. A scaling analysis 
is confirmed by simulation results. We 
obtain two classes 
of ECMC, one in which the escape time varies algebraically with the relaxation 
parameter (as for the local Metropolis algorithm) and another in which the 
escape time scales as the logarithm of the relaxation parameter. We discuss the 
connectivity of the 
hard-disk sample space, the ergodicity of local MCMC algorithms, as well as 
the meaning of packings in the context of the $NPT$ ensemble.
Our work is 
accompanied by open-source, arbitrary-precision software for \BPS (in Python) 
and for straight, reflective, forward, and Newtonian ECMC (in Go).
\\

\end{abstract}


\section{Introduction}
\label{intro}

The hard-disk system is a fundamental statistical-physics model that has been 
intensely studied since 1953. Even today, only few of its properties are known 
rigorously. Numerical simulations, notably Markov-chain Monte 
Carlo~\cite{Metropolis1953} (MCMC) and event-driven molecular 
dynamics~\cite{Alder1957}, have played a particular role in its study. The 
existence of hard-disk phase transitions~\cite{Alder1962}---although never 
proven rigorously---was asserted as early as 1962. The recent identification of 
the actual transition scenario~\cite{Bernard2009} required the use of modern 
event-chain Monte Carlo (ECMC) 
algorithms~\cite{Bernard2011,Krauth2021eventchain}.

The hard-disk model has been much studied in mathematics. A fundamental 
rigorous 
result is that the densest packing of $N$ equal hard disks (for $N\to \infty$) 
arranges them in a hexagonal lattice~\cite{Fejes1940}. This densest packing is 
locally stable: no single disk can move infinitesimally in the two-dimensional 
plane. The densest packing is furthermore collectively stable: the only 
coordinated infinitesimal displacements of a subset of disks correspond to 
symmetries, as for example uniform translations that are allowed 
by periodic boundary 
conditions~\cite{Conway1999,Torquato2010RMP,Donev2004JAP}. In 1964, 
Böröczky~\cite{Boroczky1964} constructed locally stable disk packings that are 
sparse, that is, have vanishing density in the limit $N \to \infty$. However, 
these packings are not collectively stable, and coordinated infinitesimal moves 
of several disks can escape from them. 

In this work, we construct finite-$N$ \BPS in a fixed periodic box and use them 
to build initial configurations for local Markov-chain Monte Carlo (MCMC) 
algorithms, namely  the reversible hard-disk Metropolis 
algorithm~\cite{Metropolis1953,SMAC} and several 
variants~\cite{Bernard2009,Michel2020,Klement2019} of non-reversible ECMC. In 
the Metropolis algorithm, single disks are moved one by one within a given 
range 
$\delta$. A \BP is invariant under the Metropolis algorithm if it is local, 
that 
is, if $\delta$ is small enough. ECMC is by definition local. It features 
individual infinitesimal displacements of single disks, and \BPS are likewise 
invariant. We consider \EBCS that are derived from the packings by simply 
reducing the disk radii by a factor $(1-\eps)$ where $\eps \gtrsim 0$ is the 
relaxation parameter. Our scaling theory for escape times from \EBCS
predicts 
the existence of two classes of local Markov-chain algorithms. In  one class, 
escape times grow as a power law of the relaxation parameter, whereas the other 
class features only logarithmic growth. Numerical simulations confirm our 
theory. We provide open-source arbitrary-precision software 
for \BPS and for ECMC.
We discuss the 
apparent paradox that \BPS, on the one hand, render local MCMC 
non-irreducible (that is, \quot{non-ergodic}) but on the other hand do not 
invalidate their practical use. We resolve this paradox by considering the 
$NPT$ 
ensemble (where the pressure is  conserved instead of the volume). We moreover 
advocate the usefulness  of \EBCS for modeling bottlenecks in 
MCMC and propose the  
comparison of escape times from these configurations as a useful benchmark for 
real-world problems. 

This work is organized as follows. In \sect{sec:LocallyStable}, we 
construct \BPS following the original proposal~\cite{Boroczky1964} and 
a variant due to Kahle~\cite{Kahle2012}, and we analyze their 
properties. In 
\sect{sec:MCMCEpsilonRelaxed}, we discuss local MCMC algorithms and present 
analytical and numerical results for the escape times from the \EBCS. In 
\sect{sec:Discussion}, we analyze algorithms and their escape times and discuss 
fundamental aspects, among them  irreducibility, statistical ensembles, the 
question of bottlenecks, and the difference between local and  non-local MCMC 
methods. In the conclusion (\sect{sec:Conclusion}), we point to several 
extensions and place our findings into the wider context of equilibrium 
statistical mechanics, the physics of glasses and the mechanics of granular 
materials. Our open-source arbitrary-precision software for \BPS and for ECMC 
is 
presented in \app{app:BigBoro}. 

\section{\BPS}
\label{sec:LocallyStable}

In the present section, we consider packings of  $N$ disks of radius $\sigma=1$ 
in a periodic square box  of sides $L$. The density $\eta$ is the ratio of the 
disk areas to that of the box: \begin{equation} \eta = N \pi \sigma^2 / L^2 . 
\label{equ:density} \end{equation} For concreteness, the central simulation box 
ranges from $-L/2$ to $L/2$ in both the $x$ and the $y$ direction. The periodic 
boundary conditions map the central simulation box onto an infinite hard-disk 
system with 
periodically repeated boxes or, equivalently, onto a torus. In a \LSP, each 
disk 
is blocked---at a distance $2 \sigma$---by at least three other disks (taking 
into account periodic boundary conditions), with the contacts not all in the 
same half-plane. The opening angle of a disk $i$, the largest angle formed 
by 
the contacts to its neighbors, is then always smaller than $\pi$. The maximum 
opening angle is the largest 
of the $N$ opening angle of all disks. Clearly, the packing cannot be escaped 
from through the infinitesimal single-disk 
moves of ECMC or, in Metropolis MCMC, through steps of small enough range (see 
\subsect{subsec:MCMCDescription}).

\subsection{Construction of \BPS}
\label{subsec:ConstructionBoro}

In the central simulation box, a finite-$N$ \BP is built on a 
central core placed around $(0,0)$. This core connects to four of its periodic 
copies centered at $(L,0)$, $(0, L)$, $(-L,0)$, and $(0,-L)$ by branches that 
have $k$ separate layers. A \BP shares the symmetries of the central simulation 
box. 
Cores with  different shapes, as for example that of a triangle, yield \BPS 
in other 
geometries~\cite{Boroczky1964,Pach2008,Kahle2012}.

\subsubsection{\BCORE, \KCORE}
\label{subsubsec:core}

\begin{figure}
\centering
\includegraphics{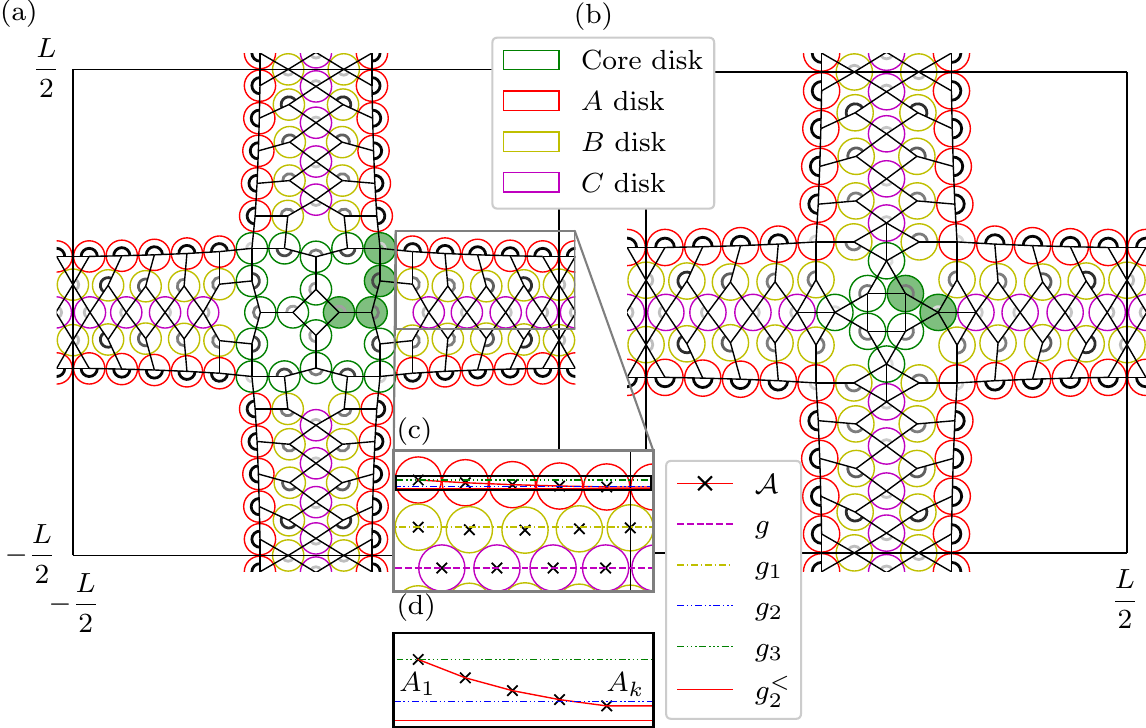}\hspace{0cm}
\caption{Hard-disk \BPS, composed of  a core and of four branches with $k=5$ 
layers, with contact graphs and  highlighted opening angles.
\subcap{a} Packing with the \BCORE~\cite{Boroczky1964}. \subcap{b} 
Packing with the \KCORE~\cite{Kahle2012}. 
\subcap{c} Detail of a branch. \subcap{d} 
Convex polygonal chain $\ACALGEO$, and horizontal lines $ g_2^<$, $g_2$, and 
$g_3$.  }
\label{fig:Boro_angles}      
\end{figure}

In the \texttt{BigBoro} software package (see \app{app:BigBoro}), we consider 
two cores. The \BCORE~\cite{Boroczky1964} consists of $20$ disks (see 
\subfig{fig:Boro_angles}{a}). Using reflection symmetry about coordinate axes 
and diagonals, this core can be constructed from four disks at coordinates 
$(\sqrt{2}, 0)$, $(2 + \sqrt{2}, 0)$, $(2 + \sqrt{6} / 2 + 1 / \sqrt{2}, 
\sqrt{6} / 2 + 1 / \sqrt{2})$, and $(2 + \sqrt{6} / 2 + 1 / \sqrt{2}, 2 + 
\sqrt{6} / 2 + 1 / \sqrt{2})$ (see highlighted disks in 
\subfig{fig:Boro_angles}{a}). The \KCORE~\cite{Kahle2012}, with a total of $8$ 
disks, is constructed from two disks at coordinates $(1, 1)$, and $(1 + 
\sqrt{3}, 0)$, using the same symmetries (see highlighted disks in 
\subfig{fig:Boro_angles}{b}). The \BCORE is locally stable if repeated 
periodically in a central simulation box that fully encloses the core disks,  
with $L/2= 3+ 
\sqrt{6}/2 + 1/\sqrt{2}$. The \KCORE is collectively stable if the outer-disk 
centers are placed at the box boundaries, with $L/2=1+\sqrt{3}$. It 
is locally stable if the outer disks are fully enclosed in the central 
simulation box, with 
$L/2=2+\sqrt{3}$.

\subsubsection{Branches---infinite-layer case (infinite $N$)}
\label{subsec:BranchesInfinite}

Following Ref.~\cite{Boroczky1964}, we first construct infinite branches that 
correspond to the $N \to \infty$ and $ \eta \to 0$ limits, without 
periodic boundary conditions. One  such branch is attached to each 
of 
the four sides of the central core so that all disks are locally stable. The 
horizontal branch that extends from the central core in the positive 
$x$-direction is 
symmetric about the $x$-axis. The half branch for $y \ge 0$ uses three sets of 
disks $\SET{A_1, A_2, \dots}$, $\SET{B_1, B_2, \ldots}$, and $\SET{C_1, C_2, 
\ldots}$, where $i= 1,2,\dots$ is the layer index. 

For the branch that is symmetric about the $x$-axis, the construction  relies 
on 
four horizontal lines~\cite{Boroczky1964}:
\begin{equation}
 \begin{array}{c|ccccc}
\text{horizontal line}   &\quad g\quad & \quad g_1 \quad& \quad g_2\quad & 
\quad g_3 \quad \\ \hline
\text{$y$-value}   &  0  & \sqrt{3} & 2\sqrt{3} & \sqrt{3} + 2\\
 \end{array} \quad .
\end{equation}
The disks $A_1$ and $B_1$ are aligned in $x$ at 
heights $g_3$ and $g_1$, respectively. All $A$ disks lie on a given 
convex 
polygonal chain $\ACAL$ between $g_2$ and $g_3 $. The chain segments on 
$\mathcal{A}$ are of length $2$ so that subsequent disks $A_i$ and $A_{i+1}$ 
block each other, and the position of $A_1$ fixes all other $A$ disks. All $C$ 
disks lie on $g$, and $C_i$ blocks $B_i$ from the right (in particular, $C_1$ 
is 
placed after $B_1$). The disk $B_i$, for $i>1$, lies between $g$ and $g_1$ and 
it blocks disks $A_i$ and $C_{i-1}$ from the right. With the position of $g_2$, 
the branch approaches a hexagonal packing for $i\rightarrow 
\infty$. After reflection about the $x$-axis,  all disks except $A_1$ and $B_1$ 
are locally stable in the infinite branch.

The \BP is completed by attaching the four branches along the four coordinate 
axes. For the \BCORE, both $A_1$ and $B_1$ are blocked by core disks (see 
\subfig{fig:Boro_angles}{a}). For the \KCORE, $B_1$ is blocked by a core disk, 
and $A_1$ is locally stable as it also belongs to another branch (see 
\subfig{fig:Boro_angles}{b}).

\subsubsection{Branches---finite-layer case (finite $N$), periodic boundary 
conditions}
\label{subsubsec:branches}

Branches can also be constructed for periodic simulation boxes,
with a finite number $k$ of layers and finite 
$N$ (see~\cite{Boroczky1964}). The branch that connects the  central core 
placed 
around $(0,0)$ with its periodic image around $(L,0)$ is then again symmetric 
about the $x$-axis but, in addition, also about the boundary of the central 
simulation box at $x=L/2$. We describe the construction of the half-branch (for 
$y 
\ge 0$) up to this boundary (see \fig{fig:Boro_angles}).

For half-branches with a finite number of layers $k$ and a finite number of 
disks $\SET{A_1 \TO A_k}$, $\SET{B_1 \TO B_k}$, and  $\SET{C_1 \TO C_{k-1}}$ 
(with their corresponding mirror images), the convex polygonal chain 
$\mathcal{A}$ lies between $g_2^<$ and $g_3$ where $g_2^<$ is an auxiliary 
horizontal line placed slightly below $g_2$ (see~\cite{Boroczky1964}). The 
horizontal lines $g$ and $g_1$ and the algorithm for placing the disks are as 
in 
\subsect{subsec:BranchesInfinite} (see \subfig{fig:Boro_angles}{c and d}). By 
varying the distance between $g_2$ and $g_2^<$, one can make disk $B_k$ 
satisfy the
additional 
requirement $x_{B_k} = x_{A_k} +1$. The position of $B_k$ then 
fixes the 
boundary of the square box ($x_{B_k} = L/2$) and $B_k$ blocks $A_k$ as well as 
the mirror image $A_{k+1}$ of $A_k$ (see \subfig{fig:Boro_angles}{c} again).

\subsection{Properties of \BPS}
\label{subsec:CommentsBoro}

The \texttt{BigBoro} software package (see \app{app:BigBoro}) implements 
two convex polygonal chains that we now discuss. It also determines the 
collective escape modes from a \BP, the space of which we also discuss.

\subsubsection{Convex polygonal chains (geometric, circular)}
\label{subsubsec:specific}

In the convex geometric chain $\ACALGEO$, the 
disks $A_i$ approach the line $g_2^<$ exponentially in $i$. In contrast,  
in the circular chain $\ACALCIRC$, all $A$ disks lie on a circle 
(including their mirror images after reflection about $x = L/2$) so that their 
opening angles are all the same.

For the convex polygonal chain $\ACAL^{\text{geo}}$, the distance between $A_i$ 
and $g_2^<$ follows a geometric progression: 
\begin{equation} 
\DIST{A_{i + 
1}}{g_2^<} = \attenuation\; \DIST{A_i}{g_2^<}, \quad \attenuation \in (0,1), 
\label{equ:RecursionA_i} 
\end{equation} 
with the attenuation parameter $\attenuation$. 
(For a horizontal branch, the distances in \eq{equ:RecursionA_i} are simply the 
difference between $y$-values.) The densities $\eta_{\text{Bör}}$ and 
$\eta_{\text{Kahle}}$ vary with $\attenuation$, and they decrease as $\sim 
1/k$ 
for large $k$ (see \tab{tab:BC_Densities}). The geometric sequence for $A_i$ 
induces that the largest opening angle, usually the one between $A_{k-1},A_k, 
A_{k+1}$, 
approaches the angle $\pi$ as $\theta_k 
=\attenuation^{k-2}(1-\attenuation)(g_3 - g_2^<)/2 
\sim \attenuation^k$, that is, exponentially in $k$ and in $L$.

\begin{table}
\centering
\begin{tabular}{c|cccc}
layers $k$ & density $\eta_{\text{Bör} }$ & density $\eta_{\text{Kahle}}$ & 
\!\!\!\!def. angle$^{\text{circ}}$ \!\!\!\! & \! \! \!\! def.
angle$^{\text{geo}}$ \!\!\!\!\\   
\hline
5    & \DENSITY{0.3957}{3.1}{-4} & \DENSITY{0.4660}{4.3}{-4} & \fpn{8.3} 
{-1} & \fpn{3.8}{-1}\\
6    & \DENSITY{0.3625}{2.9}{-4} & \DENSITY{0.4204}{3.9}{-4} & \fpn{5.3} 
{-1} & \fpn{2.5}{-1}\\
7    & \DENSITY{0.3338}{2.6}{-4} & \DENSITY{0.3820}{3.3}{-4} & \fpn{3.8} 
{-1} & \fpn{1.8}{-1}\\
8    & \DENSITY{0.3089}{2.2}{-4} & \DENSITY{0.3496}{2.8}{-4} & \fpn{2.8}
{-1} & \fpn{1.3}{-1}\\
9    & \DENSITY{0.2873}{1.9}{-4} & \DENSITY{0.3219}{2.4}{-4} & \fpn{2.2} 
{-1} & \fpn{9.9}{-2}\\
10   & \DENSITY{0.2683}{1.7}{-4} & \DENSITY{0.2982}{2.1}{-4} & \fpn{1.7} 
{-1} & \fpn{7.6}{-2}\\
15   & \DENSITY{0.2010}{9.5}{-5} & \DENSITY{0.2171}{1.1}{-4} & \fpn{7.3} 
{-2} & \fpn{2.2}{-2}\\
20   & \DENSITY{0.1604}{6.0}{-5} & \DENSITY{0.1704}{6.7}{-5} & \fpn{4.1} 
{-2} & \fpn{7.0}{-3}\\
30   & \DENSITY{0.1141}{3.0}{-5} & \DENSITY{0.1190}{3.2}{-5} & \fpn{1.8} 
{-2} & \fpn{7.4}{-4}\\
50   & \DENSITY{0.0722}{1.2}{-5} & \DENSITY{0.0741}{1.2}{-5} & \fpn{6.3} 
{-3} & \fpn{8.5}{-6}\\
100   & \DENSITY{0.0376}{3.1}{-6} & \DENSITY{0.0381}{3.2}{-6} & \fpn{1.6} 
{-3} & \fpn{1.2}{-10}\\
1000   & \DENSITY{0.0039}{3.3}{-8} & \DENSITY{0.0039}{3.3}{-8} & \fpn{1.5} 
{-5} & \fpn{7.4}{-98}\\
\end{tabular}
\caption{Parameters of \BPS for different numbers $k$ of layers.
Density window for the Böröczky and Kahle cores with 
$\ACALGEO$, obtained from $\attenuation$ between $0.0001$ and 
$0.9$. Deficit angle with respect to 
$180^\circ$ of the maximum opening angle (in degrees, same for both cores)
for $\ACALCIRC$ and for $\ACALGEO$
with $\attenuation = 0.8$. }
\label{tab:BC_Densities}
\end{table}

For the convex polygonal chain $\ACALCIRC$, all $A$ disks  lie on a
circle of radius $R$, and in particular $A_1$, which by construction 
is on $g_3$  (see \subsect{subsec:BranchesInfinite}). The circle is tangent to 
$g_2^<$ at $x=L/2$. The center of the circle lies on the vertical line at $x = 
L/2$. It follows from basic trigonometry that for large $k$, the radius of 
the circle $R$ 
scales as  $\sim k^2$ and the opening angles approach the angle $\pi$ as $ \sim
k^{-2}$.

\subsubsection{Contact graphs: local and collective stability}
\label{subsect:ContactGraphs}

The contact graph of a \BP connects any two disks whose pair distance equals 
$2$ 
(possibly accounting for periodic boundary conditions, see 
\fig{fig:Boro_angles}). In a \BP with $k \ge 1$ layers, the number $N$ of disks 
and the number $N_{\text{contact}}$ of contacts are as follows:
\begin{equation}
 \begin{array}{l|ccc}
 & \quad N \quad  & N_{\text{contact}} 
\quad\\ \hline
\text{\BCORE} & \quad 20 k + 12 & \quad 32k+20  \\
\text{\KCORE} & \quad 20 k -4 & \quad 32 k + 4 
 \end{array}\quad .
\label{equ:contacts}
\end{equation}
For all values of $k \ge 1$, the number of contacts is 
smaller than $2N-2$. 
This implies that collective infinitesimal two-dimensional 
displacements, with $2N-2$ degrees of freedom (the values of the displacements 
in $x$ and in $y$ for each disk avoiding trivial translations), can 
escape from a \BP, which is thus not 
collectively stable (see for example~\cite{Kahle2012}). 

\begin{figure}
\centering
\includegraphics{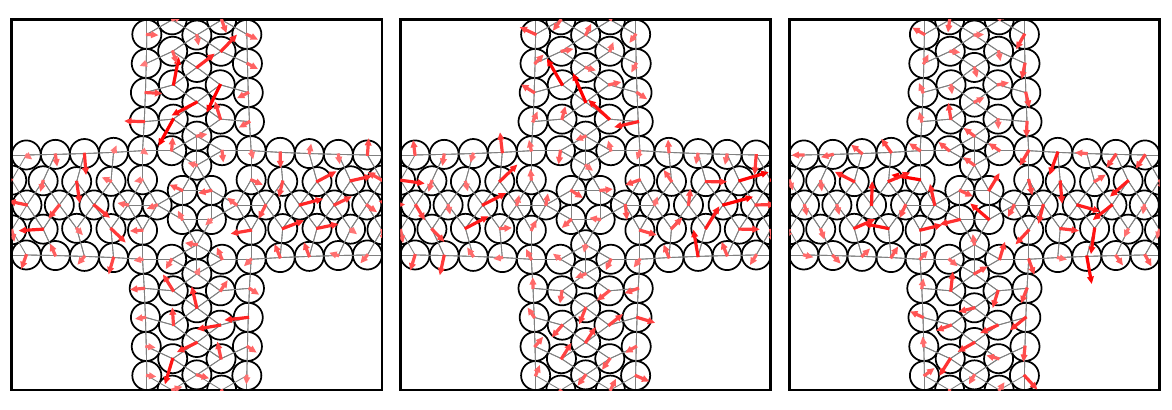}
\caption{Three of the $28$ orthogonal collective escape modes $\Deltavec$
(represented as red arrows) for the $k=5$ \BP 
with the Kahle core and $\ACALGEO$ with $\attenuation=0.7$.}
\label{fig:Escape}      
\end{figure}

When all disks $i$, at positions $\xvec_i$, are moved to $\xvec_i 
+\Deltavec_i $, the squared separation 
between disks $i$ and $j$ changes from
$| \xvec_i - \xvec_j| ^2$ to
\begin{equation}
| \xvec_i + \Deltavec_i - (\xvec_j + \Deltavec_j) | ^2  = 
| \xvec_i - \xvec_j|^2   + 
\underbrace{2 (\xvec_i-\xvec_j)\cdot(\Deltavec_i - 
\Deltavec_j)}_{\text{first-order term}}
+ |\Deltavec_i -  \Deltavec_j | ^2  
\label{equ:FirstOrderCondition}
\end{equation}
(possibly accounting for periodic boundary conditions). If the  first-order 
term 
in \eq{equ:FirstOrderCondition} vanishes for all $i$ and $j$,  the separation 
between disks in contact cannot decrease. It then increases---to second order 
in the displacements
$\Deltavec_i, \Deltavec_j$---if $\Deltavec_i \ne \Deltavec_j$, so that contact 
is lost.
The first-order term writes 
as a product of twice an \quot{escape matrix} $\MESCAPE$ of 
dimensions $N_{\text{contacts}} \times 2N$ with a $2N$-dimensional vector 
$\Deltavec = \SET{\Deltavec_1 \TO \Deltavec_N} = \SET{\Delta^x_1, \Delta^y_1, 
\Delta^x_2, 
\Delta^y_2, \dots}$. The row $r$ of $\MESCAPE$ corresponding to 
the 
contact between $i$ and $j$ has the following four non-zero entries
\begin{equation}
\begin{aligned}
\MESCAPE_{r, 2i-1}&=  x_i - x_j, \\
\MESCAPE_{r, 2i\hphantom{-1}}&=  y_i - y_j, \\
\MESCAPE_{r, 2j-1}&=-(x_i - x_j), \\
\MESCAPE_{r, 2j\hphantom{-1}}&=-(y_i - y_j) .
\end{aligned}
\label{equ:EscapeMatrix}
\end{equation}
The \texttt{BigBoro} software package (see \app{app:BigBoro}) solves for 
\begin{equation}
\MESCAPE \Deltavec = 0
\label{equ:EscapeMatrixEquation}
\end{equation}
using singular-value decomposition. For the $k=5$ \BP with the \KCORE, we find 
$28$ vanishing singular values, that comprise the two uniform 
translations. 
It follows from \eq{equ:contacts} that, because of  $28 = 2N - 
N_{\text{contact}}$, 
all contacts are linearly 
independent. The corresponding space of all collective escape modes is
$28$-dimensional (see \fig{fig:Escape}).
 
For all values of $k \ge 1$, the number of contacts in \eq{equ:contacts} is 
also larger than $N-1$. \BPS are thus collectively stable 
for displacements $\Deltavec_i$ that are constrained to a single direction (as 
for example the $x$ or $y$ direction). This impacts the behavior of 
unidirectional MCMC algorithms (see  
\subsecttwo{subsec:ECMCDescription}{subsec:EscapeTimesScalingTheory}).

\subsubsection{Dimension of the space of \BPS}

As discussed in \subsect{subsect:ContactGraphs}, each  \BP has a contact graph. 
Conversely, a given contact graph may describe \BPS 
for a continuous 
range of densities $\eta$. As an example, changing the attenuation parameter 
$\attenuation$ of the convex polygonal chain $\ACALGEO$ in 
\eq{equ:RecursionA_i} 
continuously 
moves all branch disks, and in particular disk $B_{k}$ and, therefore, the 
value 
of $L$ and the density $\eta$ (see \tab{tab:BC_Densities} for density 
windows that can be obtained in this way). We expect that locally stable 
packings exist for any density at large enough $N$. 

Moreover, the space $\BCAL$ of locally stable packings of $N$ disks of radius 
$\sigma$ in a given central simulation box is of lower dimension than the 
sample space $\Omega$: 
For each contact graph, each independent edge decreases the dimensionality by 
one. In addition there is only a
finite number of contact graphs for a given $N$. The low dimension of 
$\BCAL$ also checks with the fact that 
any packing, and more generally, any configuration with
contacts, has effectively infinite pressure (see the detailed 
discussion in \subsect{subsec:PressureEnsemble}). As the ensemble-averaged 
pressure is finite (except for the densest packing), the packings must be of
lower dimension. We conjecture $\Omega \setminus \BCAL$ to be connected for a 
given 
$\eta$ below the densest packing at large enough $N$ although, in our 
understanding, this is proven only for $\eta \sim 1/\sqrt{N}$ 
(see~\cite{DiaconisLebeauMichel2011,Baryshnikov2014}).

\section{MCMC algorithms and \EBCS}
\label{sec:MCMCEpsilonRelaxed}

In this section, we first introduce to a number of local MCMC 
algorithms 
(see \subsect{subsec:MCMCDescription}). In \subsect{sec:EscapeTimes}, we then 
determine the escape times (in the number of trials or events) after which these
algorithms escape from \EBCS, that is, from \BPS with disk radii multiplied by 
a factor $(1-\eps)$. A scaling theory establishes the existence of two classes 
of MCMC algorithms, one in which the escape time from an \EBC scales 
algebraically with $\eps$, and the other in which the scaling is logarithmic. 
Numerical simulations confirm the theory.

\subsection{Local hard-disk MCMC algorithms}
\label{subsec:MCMCDescription}

We define the reversible Metropolis algorithm with two displacement sets, from 
which the trial moves are uniformly sampled. We also consider variants of 
the non-reversible ECMC algorithm that only differ in their treatment of 
events, that is, of disk collisions. An arbitrary-precision implementation of 
the discussd ECMC algorithms (in the Go programming language) is contained in 
the \texttt{BigBoro} software package (see \app{app:BigBoro}).

\subsubsection{Local Metropolis algorithm: displacement sets}
\label{subsec:MetropolisDescription}

\begin{figure}
\centering
\includegraphics{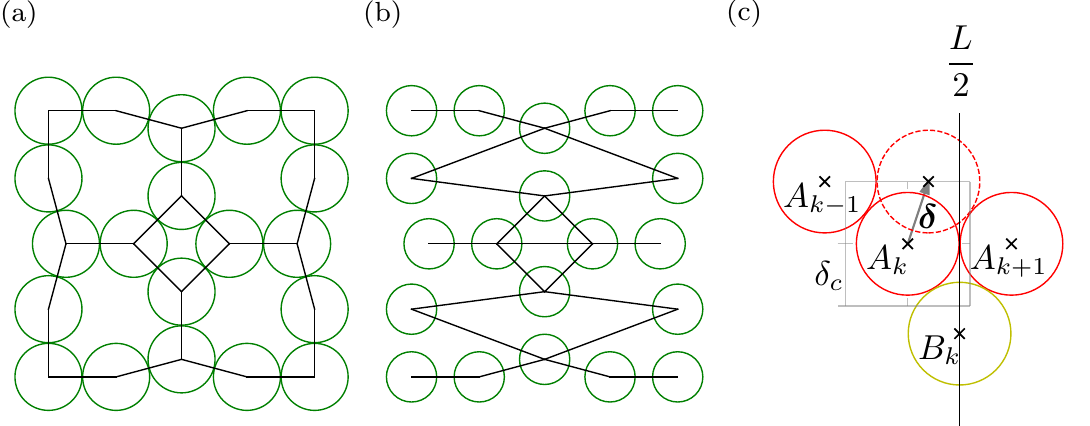}
\caption{Contact graphs, constraint graphs and minimal escape ranges. 
\subcap{a} Contact graph for a packing consisting in the \BCORE. \subcap{b} 
Constraint graph in $x$-direction for an \EBC derived from the same 
packing with $\eps = 0.25$. The edges indicate all possible 
collisions of 
straight ECMC in $x$-direction. \subcap{c} Escape move $\deltavec$ and minimal 
escape range $\delta_c$ of the Metropolis algorithm with a square-shaped 
displacement set. }
\label{fig:Constraint_graphs}
\end{figure}

The $N$ disks are at a 
position $\xvec= \SET{\xvec_1 \TO \xvec_N}$. In the local Metropolis 
algorithm~\cite{Metropolis1953},
at each time 
$t=1,2,\dots$, a trial move is proposed for a randomly chosen disk 
$i$, 
from its position $\xvec_i$ to $\xvec_i + \Delta \xvec_i$. If the trial 
produces 
an overlap, disk $i$ stays put and $\xvec$ remains unchanged. We study two 
sets for the trial moves. 
For the cross-shaped displacement set,  the trial moves are uniformly 
sampled  
within a  range $\delta$ along the coordinate axes, that is, either along the 
$x$-axis ($\Delta \xvec_i = (\ranb{-\delta,\delta}, 0)$) or along the $y$-axis 
($\Delta 
\xvec_i = (0, \ranb{-\delta,\delta})$). Alternatively, for the 
square-shaped displacement set, the trial  moves are uniformly sampled 
as 
$\Delta 
\xvec_i = (\ranb{-\delta,\delta}, \ranb{-\delta, \delta})$. A \BP is invariant 
under the local Metropolis algorithm if the range $\delta $ is smaller than a 
critical range $\delta_c$. The latter is closely related to the critical 
opening 
angle (see the discussion in \subsect{subsubsec:specific} and 
\subfig{fig:Constraint_graphs}{c}). For these packings, the critical range 
vanishes for $N \to \infty$. On the other hand, for large ranges $\delta$, the 
algorithm can readily escape from the stable configuration. For $\delta=L/2$,
the Metropolis algorithm with a square-shaped displacement set
proposes a random placement  of the disk $i$ inside the central 
simulation box. This displacement set leads to a very inefficient algorithm at 
the 
densities of physical interest, but it mixes very fast for sparse systems 
(see also \subsect{subsec:NonLocal}).

\subsubsection{Hard-disk ECMC: straight, reflective, forward, 
Newtonian}
\label{subsec:ECMCDescription}

Straight ECMC~\cite{Bernard2009} is one of the two original variants of
event-chain Monte Carlo. This Markov chain evolves in (real-valued) continuous 
time $\TMCMC$, but its implementation is event-driven. The algorithm 
is organized in a sequence of \quot{chains}, each with a chain time $\tchain$, 
its intrinsic parameter. 
During each chain time, disks move with unit velocity in one given direction 
(alternatively in $+x$ or $+y$).
A randomly sampled initial disk thus moves either until the 
chain time $\tchain$ is used up, or until, at a collision event, it collides 
with another disk, which then moves in its turn, \etcp. This algorithm is 
highly 
efficient in some applications~\cite{Bernard2009,Bernard2011,Engel2013}. During 
each chain (in 
between changes of direction), any disk can collide only with three 
other 
disks or fewer~\cite{KapferPolytope2013,Li2021}. A constraint graph with 
directed edges 
may encode these relations. This constraint graph (defined for hard-disk 
configurations) takes over the role of the contact graph (defined for packings) 
(see \subfig{fig:Constraint_graphs}{a and b}). As the moves in a chain are all 
in the same direction, the straight ECMC has only $N -1 $ degrees of freedom, 
fewer than there are edges in the constraint graph. It is for this reason 
that it may encounter the rigidity problems evoked in 
\subsect{subsect:ContactGraphs}.

In reflective ECMC~\cite{Bernard2009}, in between events, disks move in 
straight lines just as in straight ECMC. At a collision event, the target disk 
does not continue in the same direction as the active disk. Rather, the 
target-disk direction is the original active-disk direction reflected from the 
line connecting the two disk centers at contact (see~\cite{Bernard2009}). As 
all 
ECMC variants, reflective ECMC satisfies the global-balance condition. Because 
the number of disks is large,  we need not implement resamplings as is necessary 
in low-dimensional systems in order to enforce 
irreducibility~\cite{BouchardCote2018,HoellmerDipoleMany2021}. In consequence, 
the reflective ECMC has no intrinsic parameter. A variant of reflective ECMC, 
obtuse ECMC~\cite{Klement2019}, has shown interesting behavior. 

Forward ECMC~\cite{Michel2020}, of which we implement the \quot{Forward All Ref}
variant, is a family of ECMC algorithms.
After an event, the target-disk direction (of unit 
absolute 
value) is updated as follows. The component orthogonal to the line connecting 
the disks at contact is uniformly sampled between $0$ and $1$ (reflecting the 
orthogonal orientation). Its parallel component is determined so that the 
direction vector (which is also the velocity vector) is of unit norm. The 
parallel orientation remains unchanged. The forward ECMC has no intrinsic 
parameter and requires no 
resamplings. 

Newtonian ECMC~\cite{Klement2019} mimics molecular dynamics in order to 
determine the velocity of the target disk in an event. It initially samples 
disk velocities from the two-dimensional Maxwell distribution. However, at each 
moment, only a single disk is actually moving with constant velocity. At a 
collision event, the velocities of the colliding disks are updated according to 
Newton's law of elastic collisions for hard disks of equal masses, but only the 
target disks actually moves after the event. In this algorithm, the velocity 
(which indexes the Monte-Carlo time) generally differs from unity. We do not 
implement resamplings, although (like reflective ECMC) Newtonian 
ECMC is not always  
irreducible without them~\cite{HoellmerDipoleMany2021}. As in 
earlier studies for three-dimensional hard-sphere systems~\cite{Klement2019}, 
Newtonian ECMC is typically very fast for \EBCS. However, it suffers 
from frequent gridlocks (see \subsect{subsec:Gridlock}).

\subsection{Escape times from \EBCS}
\label{sec:EscapeTimes}

The principal figure of merit for a Markov chain is its mixing 
time~\cite{Levin2008}, the number of steps  it takes from the worst-case
initial condition to approach the stationary probability distribution to some 
precision level. \BPS are invariant under local Metropolis dynamics (of 
sufficiently small range) as well as under ECMC dynamics, so that the mixing 
times are, strictly speaking, infinite. Although they cannot be escaped from, 
the packings make up only a set of measure zero in sample space, and might thus 
be judged irrelevant. 

However, as we will discuss in the present section, the situation is more 
complex. For every \BP, an associated \EBC keeps the central simulation box and 
the disk positions, but reduces the disk radii from $1$ to $1-\eps$. An \EBC 
effectively defines a finite portion of configuration space (the spheres of 
radius $\eps$ around each disk position, see \sect{subsec:ScalingArguments}). 
All MCMC algorithms considered in this work
escape from 
these configurations in an escape times that diverges as $\eps \rightarrow 0$. 
We suggest that escape times are analogous to mixing times. In consequence, a 
finite portion of sample space is confined on times larger than an arbitrary 
constant. We suggest that the substantial differences between escape times may 
be relevant for real-world applications. The divergence of escape times for 
$\eps \to 0$ is specific to the $NVT$ ensemble (see 
\subsect{subsec:PressureEnsemble}).

\subsubsection{Nearest-neighbor distances and escape times}

In a \BP, disks are locally stable, and they all have a nearest-neighbor 
distance of $2$. The packings are sparse, and the nearest-neighbor distance is 
thus smaller than its $\sim 1/ \sqrt{\eta}$ equilibrium value. To track the 
escape from an \EBC, we monitor the maximum nearest-neighbor 
distance:
\begin{equation}
\MAXNEAR(\TIME) = \max_i \glc \min_{j (\neq i)} |\xvec_{ij}(\TIME)| 
\grc,
\label{equ:MaxNear}
\end{equation}
where $|\xvec_{ij}(\TIME)| = |\xvec_j(\TIME) - \xvec_i(\TIME)|$ is the
distance between disks $i$ and $j$ (possibly corrected 
for periodic boundary conditions). For the Metropolis algorithm, we compute 
$d(t)$ once every $N$ trials, and $t$ denotes the 
integer-valued number of 
individual 
trial moves. For ECMC, we sample $d(t)$ and the number of events in intervals 
of 
the sampling time. In \eq{equ:MaxNear}, $t$ then denotes the integer-valued 
number of events. Starting from an \EBC, $d(t)$ typically remains at $d(t) \sim 
2 + 
\bigOb{\eps}$ for a long time until it approaches the equilibrium value in a 
way 
that depends on the algorithm. We define the escape time $\ESCAPE$, an integer, 
as the  time $t$ at which $d(t)$ has increased  by ten percent:
\begin{equation}
\ESCAPE = \minc{ \TIME : \MAXNEAR(\TIME) > 2  (1+ 
\MAXNEARBOUND)},
\label{equ:escape}
\end{equation}
with $\MAXNEARBOUND=0.1$. 
Our conclusions are robust with respect to the value of $\MAXNEARBOUND$.

\subsubsection{Escape times---scaling theory}
\label{subsec:EscapeTimesScalingTheory}

The local Metropolis algorithm and the straight ECMC both have an intrinsic 
parameter, namely the range $\delta$ of the displacement set or the chain time 
$\tchain$. 
These two parameters play a similar role.

Two limiting cases can be analyzed. For the Metropolis algorithm at small 
$\delta$, a trajectory spanning a constant distance is required to escape from 
an 
\EBC. As the dynamics is diffusive, we have $\const = \delta \sqrt{\ESCAPE}$. 
For the straight ECMC with small chain times $\tchain$, the effective 
dynamics (after subtraction of the uniform displacement), is again diffusive. 
This leads to: 
\begin{equation}
 \ESCAPE(\delta) \sim 
 \begin{cases}
 \const / \delta ^ 2  & \text{(Metropolis)},\\
 \const / \tchain ^ 2  & \text{(ECMC)},
 \end{cases}\quad
 \text{(for small $\delta $, $\tchain$)}.
\label{equ:EscapeSmallDelta}
\end{equation}
On the other hand, even for large $\delta$ or $\tchain$, the Markov chain must 
make a certain number of moves on a length scale $\eps$ in order to escape from 
the \EBC. In the Metropolis algorithm, the probability for a trial on this 
scale 
is $\eps / \delta$ for the cross-shaped displacement set, and $\eps^2 
/\delta^2$ for 
the 
square-shaped displacement set. For the straight ECMC with large 
$\tchain$, all 
displacements beyond a time $\sim \eps$ (or, possibly, $\sim N \eps$) 
effectively cancel 
each other, because the constraint graph is rigid. This leads to: 
\begin{equation}
 \ESCAPE(\delta) \sim 
 \begin{cases}
  \delta^2 / \eps^2 & \text{(Metropolis---square)},\\
  \delta / \eps & \text{(Metropolis---cross)},\\
  \tchain / \eps & \text{(straight ECMC)},\\
 \end{cases} \quad \text{(for large $\delta$, $\tchain$).}
\label{equ:EscapeLargeDelta}
\end{equation}
The two asymptotes of \eqtwo{equ:EscapeSmallDelta}{equ:EscapeLargeDelta} form a 
\quot{$V$} with a base $\delta_V$ at $\sim \sqrt[3]{\eps}$ (for 
the 
Metropolis algorithm with a cross-shaped displacement set, and for straight 
ECMC) and  
at $ \sim \sqrt{\eps}$ (for the Metropolis algorithm with a square-shaped move 
set). The resulting optimum, the  minimal  escape time with respect to 
$\eps$, is
\begin{equation}
 \ESCAPE  \simeq  \ESCAPE(\delta_V)
\sim 
 \begin{cases}
    \eps^{-1} & \text{(Metropolis---square)}, \\
   \eps^{-2/3} & \text{(Metropolis---cross)},\\
    \eps^{-2/3} & \text{(straight ECMC)}. \\
 \end{cases}
\label{equ:EscapeMinDeltaMetropolis}
\end{equation}
These scalings  balance two requirements: to move by a constant distance (which 
favors large $\delta$ or $\tchain$) and to move on the scale $\eps$ (which 
favors 
small $\delta$ or $\tchain$).

The forward, reflective, and Newtonian ECMC move in any direction, even in the 
absence of resamplings, so that their displacement sets are  $2N$-dimensional. 
This avoids the rigidity problem of straight ECMC (the fact that the number of 
constraints can be larger than the number of degrees of 
freedom). These algorithms introduce  no intrinsic scale (as $\delta$ or 
$\tchain$). 
The effective step size of moves may thus  adapt as the configuration gradually 
escapes from the \EBC. The step size is 
initially on the scale $\eps$, but then grows on average by a constant 
factor at each event, reaching a scale $\eps' > \eps$ after a time $\sim \ln ( 
\eps'/\eps)$. The scale $\eps'$ at which the algorithms break free is 
independent of the initial scale $\eps$, and we expect a 
logarithmic scaling of the escape time (measured in events):
\begin{equation}
 \ESCAPE \sim \ln(1 / \eps)\quad
\text{(reflective, forward, and Newtonian ECMC)}.
\label{equ:EscapeMinDeltaForward}
\end{equation}

\begin{figure}[htb]
\centering
\includegraphics{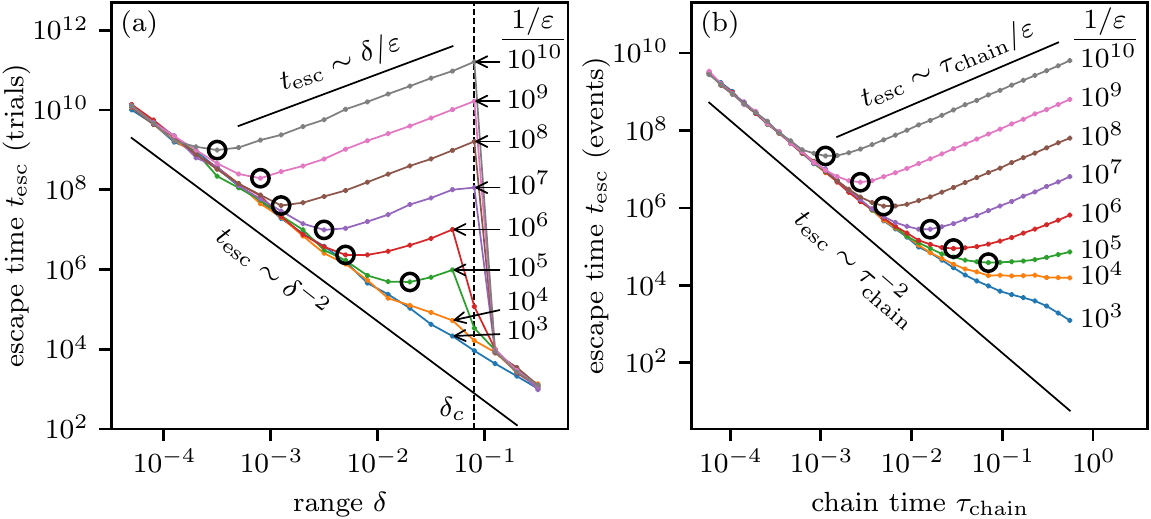}
      \caption{ Escape times from the $k=5$ \EBC (\KCORE, 
      $\ACALGEO$, $\attenuation=0.7$) for different 
      $\eps$.  \subcap{a} $\ESCAPE$ (in trials) \emph{vs.} range 
      $\delta$ for the cross-shaped displacement set.
      \subcap{b} Escape time $\ESCAPE$ (in events) \emph{vs.} chain time 
      $\tchain$
      for straight ECMC. 
      Asymptotes are from \eqtwo{equ:EscapeSmallDelta}{equ:EscapeLargeDelta}. }
       \label{fig:rev_Boro_sweep}
\end{figure}

\subsubsection{Escape times---computation results}
\label{subsec:EscapeTimesComputation}

We now test the scaling theory  (see \subsect{subsec:EscapeTimesScalingTheory}) 
of the escape times for $k=5$ \EBCS for small relaxation parameter $\eps$. For 
the local Metropolis algorithm and the straight ECMC, the predicted behavior of 
$\ESCAPE$ for small and large parameters $\delta$ or $\tchain$ is clearly 
visible (see \eq{equ:EscapeMinDeltaMetropolis} and \fig{fig:rev_Boro_sweep}). 

\begin{figure}[htb]
\centering
\includegraphics{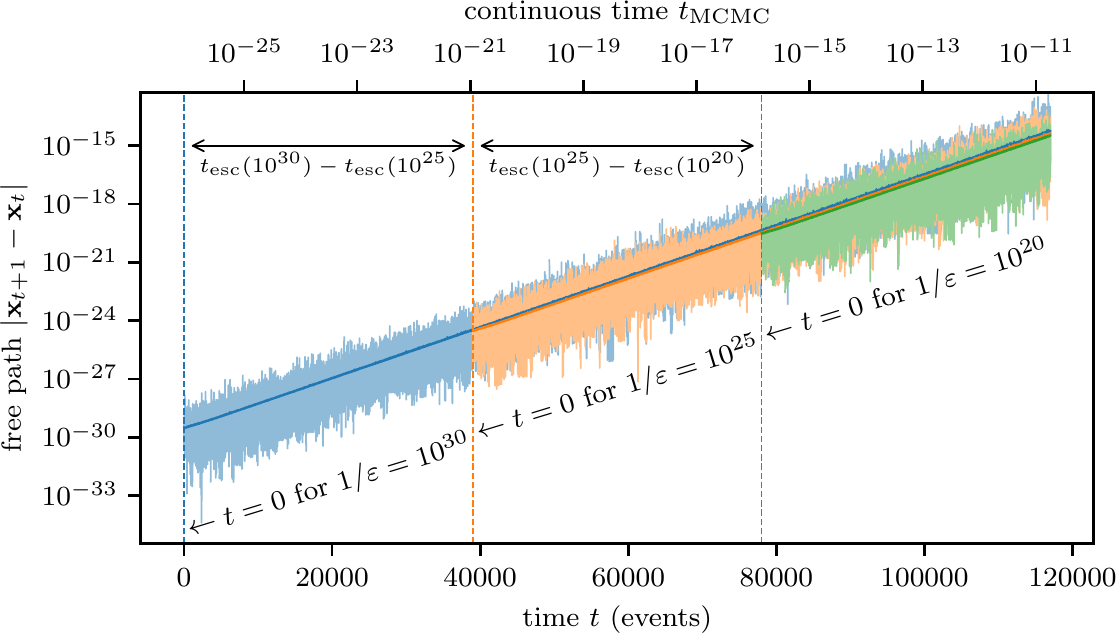}
      \caption{ Free path (equivalently: Monte-Carlo time between 
events) for the forward ECMC algorithm started from three \EBCS (\KCORE,
$\ACALGEO$, $\attenuation = 0.7$)
with $\eps = 10^{-30}$, $10^{-25}$ and $10^{-20}$. Integer time $t$ 
(lower $x$-axis)
counts  events, while $\TMCMC$ (upper $x$-axis) is the real-valued continuous 
Monte Carlo time. Event times are shifted. Expanded light curves show single 
simulations for each $\eps$, dark lines average over $10000$ simulations.}
\label{fig:MeanFreePathForward}
\end{figure}

The absence of an imposed scale for displacements manifests itself in 
the 
forward ECMC in the logarithmic dependence on time of the mean free path, that 
is, the ensemble-averaged displacement between events. As the velocity has unit 
value, the free path is equal to the difference of Monte-Carlo times $\TMCMC(t 
+1) - \TMCMC(t)$ between subsequent events. Individual evolutions as a function 
of time $t$ for 
small relaxation parameters  $\eps$ and  $\eps'$ nicely overlap when shifted 
by their escape times (see \fig{fig:MeanFreePathForward}). 
The time $t$ here refers to the number of events and not to the 
Monte-Carlo time $\TMCMC$, which depends exponentially on the number 
of events $t$. Starting from an \EBC with $\eps = 10^{-30}$, as an example, the 
same number of events is on average required to move from a mean free 
path of 
$\sim 10^{-30}$ to $10^{-25}$, as from a mean-free path $\sim 10^{-25}$ to 
$10^{-20}$ (see \fig{fig:MeanFreePathForward}). 

\begin{figure}[htb]
\centering
\includegraphics{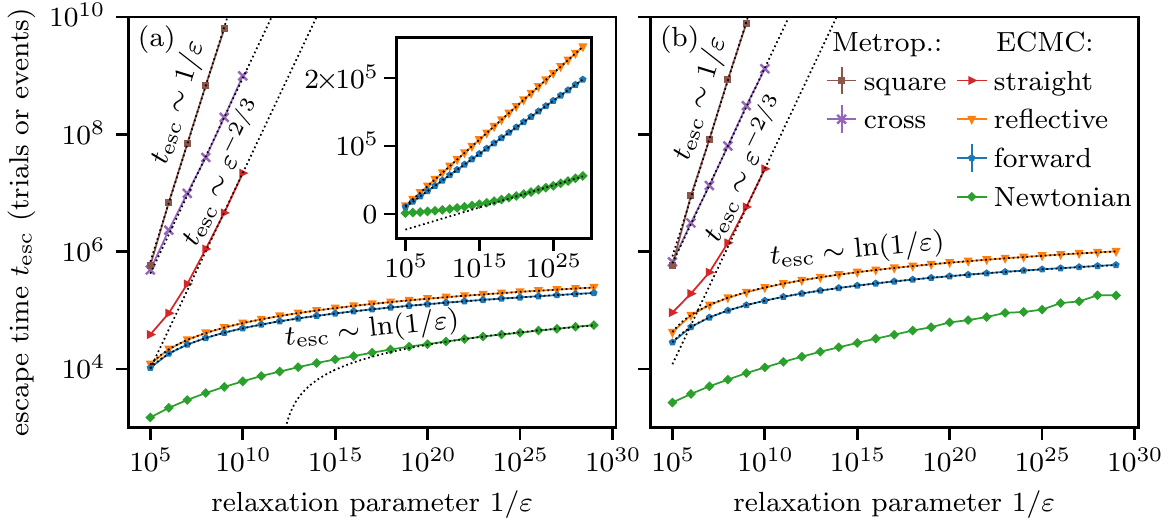}
\caption{Escape time $\ESCAPE$ from $k=5$ \EBCS ($\ACALGEO$, $\attenuation = 
0.7$) for local MCMC algorithms (where 
applicable: with optimized parameters). \subcap{a} $\ESCAPE$ (\KCORE) 
for the Metropolis 
algorithm and straight ECMC. Inset: log--lin plots suggesting logarithmic 
scaling for the forward, reflective, and Newtonian 
ECMC. 
\subcap{b} $\ESCAPE$ for the \BCORE. Newtonian ECMC has frequent gridlocks for 
small $\eps$.}
\label{fig:Escape_time}
\end{figure}

Overall, escape times (with optimized intrinsic parameters for the Metropolis 
algorithm and for straight ECMC), validate the algebraic scalings of 
\eq{equ:EscapeMinDeltaMetropolis}, on the one hand, and the logarithmic scaling 
of \eq{equ:EscapeMinDeltaForward}, on the other (see \fig{fig:Escape_time}). 
Newtonian ECMC appears \emph{a priori} as the fastest variant of ECMC. However, 
it frequently gets gridlocked, \ie, trapped in circles of repeatedly active 
disks with a diverging event rates. Gridlocks also rarely appear in
straight and reflective ECMC.  In runs that end in gridlock, escape times 
are very large, possibly diverging (in 
\figs{fig:rev_Boro_sweep}{fig:Escape_time}, median escape 
times are therefore displayed for these algorithms, rather than the means). The 
fraction of gridlocking simulations increases 
with 
$1/\eps$. For the \KCORE, this effect is negligible for all
$\eps$. For the \BCORE, Newtonian ECMC runs into 
gridlock for roughly one third of 
individual simulations for $\eps = 10^{-29}$ 
(see 
\subfig{fig:Escape_time}{b}, the logarithmic scaling is distorted even for the 
median).
For further discussion of gridlocks, see \subsect{subsec:Gridlock}.

\section{Discussion}
\label{sec:Discussion}
In  the present section, we discuss our results for the escape times 
(\subsect{subsec:DiscussionEscapeTime}), as well as a number of more 
fundamental 
aspects of \BPS in the context of MCMC (\subsect{subsec:DiscussionFundamental}).

\subsection{Analysis of measured escape times}
\label{subsec:DiscussionEscapeTime}

ECMC is a continuous-time MCMC method, and its continuous time 
$t_{\text{MCMC}}$ takes the place of the usual count of discrete-time 
Monte-Carlo trials. In ECMC, each chain corresponds to a segment 
$[t_{\text{MCMC}}, t_{\text{MCMC}} + \tchain)$ of Monte-Carlo time. However, 
ECMC is 
event-driven. The time $t$, and especially the escape time $\ESCAPE$, are
integers, and they count events. The computational effort in hard-disk ECMC is 
\bigOb{1} per event, using a cell-occupancy system that is also implemented in 
the \texttt{BigBoro} software package. In several of our algorithms, the 
times $t$ and $t_{\text{MCMC}}$ are not proportional to each other, because the 
mean-free path (roughly equivalent to the time between events) evolves during 
each individual run.

\subsubsection{Range of speedups}
\label{subsec:RangeSpeedups}

The speedup realized by lifted Markov chains, of which ECMC is a representative, 
corresponds to the transition from diffusive to ballistic 
transport~\cite{Diaconis2000,Chen1999,Krauth2021eventchain}. For Markov chains 
in a finite sample space $\Omega$, the Monte-Carlo time for mixing of the lifted 
Markov chain cannot be smaller than the square root of the mixing time for the 
original (collapsed) chain. The remarkable power-law-to-logarithm speedup in 
$\eps$ realized by some of the ECMC algorithms concerns times which  measure the 
number of events. The Monte-Carlo escape times probably conform to the 
mathematical bounds, although it is unclear how to approximate hard-disk MCMC 
for  $\eps \to 0$ through a finite Markov chain. Mathematical results for the 
escape times from locally blocked configurations would be extremely interesting, 
even for models  with a restricted number of disks.

\subsubsection{Space of \EBCS} 
\label{subsec:ScalingArguments}

Any \EBC is merely a sample in a space $\BCAL_\eps$ of volume $\sim \eps^{2N}$. 
We have in fact checked that the position $\xvec_i$ of disk $i$ in that 
configuration can be replaced by $\xvec_i + \hat{\epsvec}_i$ (where 
$\hat{\epsvec}_i$ a random vector inside the circle of radius $\eps$) without 
affecting the scaling of  escape times expressed in 
\eqtwo{equ:EscapeMinDeltaMetropolis}{equ:EscapeMinDeltaForward}. For a given 
upper limit $\tcpu$ of CPU time, this corresponds to a volume of $\BCAL_\eps$ 
(that cannot be escaped from in $\tcpu$) scaling as $\sim \tcpu^{-3 N}$, for 
example, for the straight ECMC and scaling as  $\sim  \expb{-2N \tcpu}$ for the 
forward ECMC. We expect $\BCAL_\eps$ to have a triple 
role, as a space of configurations containing  bottlenecks (see 
\subsect{subsec:Gridlock}), as a space of configurations that the Monte-Carlo 
dynamics cannot practically  escape from, but maybe also as a space that it 
cannot even access. 

The volume of \quot{practically}  stable configurations, as well as the 
corresponding 
changes in the free energy per disk are probably unmeasurably small. It is 
however remarkable that   these excluded volumes cannot be escaped 
from. In many MCMC algorithms for physical systems, as for 
example the Ising mode, parts of sample space are practically
excluded because of their low Boltzmann weight, but they do not feature
diverging  escape times at finite $N$.

\subsubsection{Gridlock of hard-disk ECMC algorithms, resamplings}
\label{subsec:Gridlock}

ECMC algorithms for soft potentials require random numbers at each event. In 
contrast, the hard-disk ECMC algorithms of 
\subsect{subsec:ECMCDescription} except of the forward ECMC treat events 
through deterministic collision rules.
At high density, this can make them susceptible to gridlock, 
in other words to diverging event rates of chains with successive disks in 
permanent contact. The Monte-Carlo time between events then goes to 
zero. Gridlock plays no role in large systems at reasonable densities, but it 
has been discussed in straight ECMC~\cite{Weigel2018}. 

Gridlock is the very essence of  ECMC dynamics that starts from a \BP, but 
it also appears as a final state for \EBCS. In runs from such 
configurations, we observe gridlock mostly for Newtonian ECMC  with the 
\BCORE, rendering the analysis of its scaling behavior with the relaxation 
parameter $\eps$ impossible. It also rarely appears in  straight and reflective 
ECMC for the smallest  $\eps$. Because of the infinite event 
rate, gridlock cannot be 
remedied through resamplings after a finite Monte-Carlo time $\TMCMC$. To 
overcome gridlock, one can probably introduce event-based randomness to 
Newtonian ECMC  as is done in forward ECMC.

\subsection{\BPS and local MCMC: fundamental aspects}
\label{subsec:DiscussionFundamental}

We now discuss fundamental aspects of \BPS, from the issue of irreducibility 
to the question of statistical ensembles, the connection with 
bottlenecks and, finally, to non-local MCMC algorithms.

\subsubsection{Irreducibility of local hard-disk MCMC}
\label{subsec:IrreducibilityLocalMCMC}

Strictly speaking, ECMC can be irreducible only if $\Omega/\BCAL$ is connected, 
where $\BCAL $ is a suitably defined space of locally stable 
configurations. Packings in $\BCAL$ (a space of low dimension) 
are certainly invariant under any version of the ECMC algorithms, so that they 
cannot evolve towards other samples in $\Omega$.
Connectivity in $\Omega / \BCAL$
would at least assure that this space can be sampled.
In addition it appears necessary to guarantee that a well-behaved 
initial configuration cannot evolve towards $\BCAL$ or even towards an 
$\eps$-environment around it. These two properties
appear not clearly established for finite 
densities $\eta$ and for large $N$. (At small $N$, 
counter-examples are easy to construct.) In other models, for example the 
Ising 
model of statistical physics, irreducibility can be proven for any $N$.

These unresolved mathematical questions concerning irreducibility do not shed 
doubt on the practical 
usefulness of MCMC for particle systems. First, the concept of local stability 
is restricted to hard disks and hard spheres (that is, to potentials that are 
either zero or infinite). The phase diagram of soft-disk models can be 
continuously connected to the hard-disk case~\cite{Kapfer2015PRL}. 
For soft disks, irreducibility is trivial, but the sampling speed of 
algorithms remains crucial. Second, in 
applications, one may change the thermodynamic ensemble. In the $NPT$ ensemble 
(further discussed in \subsect{subsec:PressureEnsemble}), the central 
simulation 
box fluctuates in size and can become arbitrarily large. In this ensemble, 
irreducibility follows from the fact that large enough  simulation boxes are 
free of steric constraints. Again, the question of mixing and correlation time 
scales is primordial. Third, practical simulations that require some degree of 
irreducibility are always performed under conditions where the simulation box 
houses a number of effectively independent copies of the system. This excludes 
the crystalline or solid phases. Monte Carlo simulations of such phases are 
more 
empirical in nature. They  require a careful choice of initial states, and are 
then not expected to visit the entire sample space during their time  
evolution. 
Fundamental quantitative results can nevertheless be 
obtained~\cite{Bolhuis1997}.

\subsubsection{\BPS and the $NPT$ ensemble}
\label{subsec:PressureEnsemble}

The concepts of packings and of local and collective stability make sense only 
in the  $NVT$ ensemble, that is,  for a constant number of particles and for a 
simulation box with fixed shape and volume (the temperature $T = 1/\beta$ that 
appears in $NVT$ plays no role in hard-disk systems~\cite{SMAC}). In the $NPT$ 
ensemble, the pressure $P$ is constant, and the size of the simulation box may 
vary. The equivalence of the two ensembles is proven~\cite{Ruelle1999} for 
large 
$N$, so that the choice of ensemble is more a question of convenience than of 
necessity. As we will see, in the $NPT$ ensemble, tiny relaxation parameters 
(as 
$\eps = 10 ^{-29}$ in \fig{fig:Escape_time}) are not maintained 
for normal pressures and system sizes. 

To change the volume at constant pressure, one may, among others, proceed to 
\quot{rift volume changes} (see~\cite[Sect. VI]{Michel2014JCP}) or else to 
homothetic transformations of the central simulation box. We discuss this 
second 
approach (see~\cite[Sect. 2.3.4]{SMAC}), where the disk positions (but not the 
radii) are rescaled by the box size $L$ as: 
\begin{equation}
\xvec = (\xvec_1 \TO\xvec_N) \to \alphavec = (\alphavec_1 \TO \alphavec_N) 
\quad \text{with 
$\alphavec_i = \xvec_i/L$}.
\end{equation}
Each configuration is then specified by an $\alphavec$ vector
in the $2N$-dimensional periodic unit square
and an associated 
volume $V = L^2$, which must satisfy $ V \ge \Vcut(\alphavec)$. A 
classic MCMC algorithm~\cite{Wood1968} directly samples the volume at fixed 
$\alphavec$ from a gamma distribution above 
$\Vcut(\alphavec)$, below which $(\alphavec, V)$ ceases to 
represent a 
valid hard-disk configuration~\cite[eq. 2.19]{SMAC}. Typical sample volumes are 
characterized by  $\beta P (V - \Vcut) \sim 1$, and with $V = 
(\Lcut + \Delta L) ^2 $, it follows that 
\begin{equation}
 \frac{\Delta L}{L} \sim \eps \sim \frac{1}{\beta P \Vcut} \quad \text{(at 
fixed $\alphavec$)}. 
\label{equ:PressureNPT} 
\end{equation}
This equation illustrates that a packing, with $\eps \to 0$, is realized as a 
typical configuration only in the limit $\beta P \to \infty$. For the \BPS of 
\fig{fig:Boro_angles}, we have $L\simeq 20$, and a typical value for the 
pressure for hard-disk systems is $\beta P \sim 1$, which results in $\eps \sim 
10^{-3}$. In the $NPT$ ensemble, as a consequence, escape times from a packing 
naturally correspond to a relaxation parameter $\eps \sim 1/ (\beta P V)$, in 
our example to $\ESCAPE(\eps \sim 10^{-3})$,  which is \bigOb{1}.

The above $NPT$ algorithm combines constant-volume $NVT$-type moves of 
$\alphavec$ with the mentioned direct-sampling moves of $V$ at fixed 
$\alphavec$. In practice, however, $NPT$ calculations are rarely performed in 
hard-disk systems~\cite{Wood1970,Lee1992}. This is because, 
as discussed in \eq{equ:PressureNPT}, 
the expected single-move displacement in volume at fixed $\alphavec$ is  
$\Delta V \sim 1/( \beta P)$, so 
that $\Delta V/V \sim 1 /N$ (because $N \sim V$ and $\beta P \sim 1$).
The 
fluctuations of the equilibrium volume $\Veq$ (averaged over $\alphavec$) scale 
as $\sqrt{\Veq}$, which implies
$\Delta \Veq/ \Veq
\sim 1 
/\sqrt{N}$. The volume-sampling algorithm requires 
$\sim N $ single
updates of the volume to go from the $1/N$ scale of volume fluctuations at 
fixed $\alphavec$ 
to the $1/\sqrt{N}$ scale of the fluctuations of $\Veq$ at equilibrium. This 
multiplies with the number of 
steps to decorrelate at a given 
volume. In practice, it has proven more successful to perform single $NVT$ 
simulations, but to restrict them  to  physical parameters where the central 
simulation box 
houses a finite number of effectively independent systems mimicking 
constant-pressure configurations.

\subsubsection{Bottlenecks in MCMC algorithms}
\label{subsec:Bottlenecks}

Markov chains 
can be interpreted in terms of a single bottleneck partitioning the sample 
space into two pieces~\cite[Sect. 7.2]{Levin2008}. The algorithmic equilibrium 
flow across the bottleneck sets the conductance of an algorithm, which 
again 
bounds mixing and correlation times. Ideally, 
MCMC algorithms would be benchmarked through their conductances.

In the hard-disk model, the 
bottleneck has not been identified, so that the benchmarking and the analysis 
of 
MCMC algorithms must rely on empirical criteria. However, \BPS and the 
related \EBCS may well exemplify possible bottlenecks and 
the escape times studied in \subsect{sec:EscapeTimes} may model mixing times. 
They certainly 
provide lower bounds. Most importantly, the benchmarks obtained by 
comparing escape times may carry important lessons on the relative merits of 
sampling
algorithms.

\subsubsection{\BPS and non-local MCMC}
\label{subsec:NonLocal}

In this work, we concentrate on local MCMC algorithms, with infinitesimal 
displacements (for ECMC) or very small displacements (for the Metropolis 
algorithm), because real-life continuous-space problems usually require the 
use of local methods~\cite{Krauth2021eventchain}.
Global-move algorithms, as the cluster 
algorithms in spin systems, rely on \emph{a priori} probabilities for 
many-particle moves that are too 
complicated. On the other hand, global single-particle moves are 
related to the 
single-particle insertion probabilities, in other words to
fugacities (the exponential 
of the negative chemical potential) that are prohibitively small.

In view of the scarcity of exact results for hard-disk MCMC algorithms, 
we 
now discuss the global-move Metropolis algorithm in which at each time 
step a randomly chosen disk is placed at a random position inside the box. This 
corresponds to the Metropolis algorithm of 
\subsect{subsec:MetropolisDescription} with a square-shaped 
displacement set and a range $\delta=L/2$. 
This non-local algorithm has no problem escaping from a \BP.
Moreover, it is proven 
to mix in \bigOb{N \loga{N}} steps at densities 
$\eta < 1/6$~\cite{Kannanrapidmixing2003,Helmuth2020} (see 
also~\cite{BernardChanalKrauth2010}). This result has made it possible to prove
that  the liquid phase in the hard-disk system extends at least to 
the density $\eta=1/6$~\cite{Helmuth2020}. 
The density bound for the algorithm (which yields a bound for the stability of 
the liquid phase) 
is much smaller than the empirical  density bound for the liquid phase, 
at $\eta \simeq 0.70$. At this higher density, the global-move 
Metropolis algorithm 
and the more general hard-disk cluster algorithm~\cite{Dress1995} are 
almost totally stuck. For applications, we imagine \BPS to be part of 
configurations at such high densities, where global moves cannot be used.

\section{Conclusion}
\label{sec:Conclusion}

Building on an early breakthrough by Böröczky, we have studied in this work
locally stable hard-disk packings. \BPS are sparse, with arbitrarily small 
densities for large numbers $N$ of disks. We constructed different types 
of these packings to arbitrary precision for finite $N$ and 
made our implementation openly accessible. \BPS are locally, but not 
collectively stable. Using singular-value decomposition (in an implementation 
that is included in our open-source software) we explicitly exposed the 
unstable 
collective modes. We furthermore reduced the radius of  \BPS slightly, and 
determined the escape times from \EBCS as a function of the parameter $\eps$ 
for 
a number of local MCMC algorithms, including several variants of ECMC, 
arbitrary-precision implementations of which are also made openly available. 
Although the algorithms depart from each other in seemingly insignificant 
details only, we witnessed widely different  escape times, ranging from 
$1/\eps$ to 
$\log(1/\eps)$. Our theory suggested that the significant speedup of 
some of the algorithms is rooted in their event-driven nature coupled to their 
lack of an intrinsic scale. 
We pointed to the 
importance of statistical ensembles to reconcile the obvious loss of 
irreducibility in the presence of \BPS with the proven practical usefulness of 
local hard-disk MCMC algorithms. 

We expect the observed differences in escape times to carry over to real-world 
ECMC implementations. In statistical mechanics, bottlenecks and escape times
possibly play an important role in polymer physics and complex molecular 
systems 
and some of the  algorithms studied here may find useful applications. 
Escape times may also play an important role in the study of glasses 
and in granular matter, where the high or even infinite pressures favor local 
configurations that resemble the mutually blocked disks in the \EBCS. We 
finally 
point out that the very concept of locally blocked packings naturally extends 
to higher dimensions.


\section*{Conflict of interest}
The authors declare that they have no conflict of interest.

\appendix

\section{\texttt{BigBoro} software package: outline, license, access}
\label{app:BigBoro}

\normalsize

The open-source \texttt{BigBoro} software package consists 
of three parts: First, the arbitrary-precision Python script 
\texttt{construct\_packing.py} 
constructs finite-$N$ \BPS of hard disks in a periodic square box. Second, 
the Python script \texttt{collective\_escape\_modes.py} computes collective 
infinitesimal displacements of hard disks in a packing that result in an 
escape. Third, the arbitrary-precision Go application
\texttt{go-hard-disks} performs 
hard-disk ECMC simulations that may start from \EBCS derived from \BPS.

\subsection{Python script \texttt{construct\_packing.py}}

The arbitrary-precision Python script \texttt{construct\_packing.py} 
implements the construction of
\BPS. It allows for 
the Böröczky or Kahle cores (see \sect{subsubsec:core}), and connects them 
to branches with a finite number of layers (see 
\sect{subsubsec:branches}). The geometric convex polygonal chain $\ACALGEO$ 
with different attenuation parameters $\attenuation$, and the circular 
$\ACALCIRC$ are implemented (see \sect{subsubsec:specific}). The core, the 
number of layers, and the convex polygonal chain are specified using 
command-line arguments that are described in the \texttt{README.md} file of the 
package, as well as in the output of the script's \texttt{--help} command-line 
option. The docstrings of the script contain further information.

The Python script \texttt{construct\_packing.py} uses 
arbitrary-precision 
decimal floating-point arithmetic (using the \texttt{decimal} module of 
Python's standard library). Two additional command-line options specify the 
number of decimal digits, and the precision $b$ of the 
bisection search for the value $g_2^<$  that renders the \BP compatible with
periodic 
boundary conditions (see \sect{subsubsec:branches}). In general, $b$
should be smaller than the number of 
places of the decimals. The script modifies $g_2^<$ until $1 \leq x_{B_k} - 
x_{A_k} \leq 1 + 10^{-b}$. If the bisection search succeeds, the script first 
tests 
that no pair of disks has a distance smaller than $2 - 
10^{-b + 
2}$. Second, it checks that  every disk has at least three contacts with 
distances in 
the interval 
$[2 - 10^{-b + 2}, 2 + 10^{-b + 2}]$, and, finally, that the total number of 
contacts agrees with \eq{equ:contacts}. The final configuration and 
its parameters (as for example the system length) are stored in a 
human-readable 
format in a specified output file.

The \texttt{example\_packings} directory of \texttt{BigBoro} contains several 
\BPS.
The packings are contained in 
corresponding subdirectories (as for 
example 
\texttt{kahle\_geometric\_5}). The headers of these files contain
the values of the command-line arguments for
\texttt{construct\_packing.py}. A plot of each 
example configuration is provided. 
The different packings in \texttt{kahle\_geometric\_5} and 
\texttt{boro\_geometric\_5} (see \fig{fig:Boro_angles}) were heavily used in 
this work.
Although the bisection 
search of the Böröczky-packing construction usually requires an increased 
precision, the high-precision packings with small enough $k$
may be used as input for standard double-precision 
applications. For simplicity and improved readability, we provide 
\texttt{packing\_double.txt} files that store the configurations with 
double precision, where applicable.

\subsection{Python script \texttt{collective\_escape\_modes.py}}

The double-precision Python script 
\texttt{collective\_escape\_modes.py} 
identifies the orthonormal basis vectors $\Deltavec_a$ of the escape matrix 
$\MESCAPE$ 
from a packing $\xvec$ (see 
\eq{equ:EscapeMatrix}) that have zero singular values. This is the solution 
space for $2N$-dimensional displacements $\Deltavec = \SET{\Delta^x_1, 
\Delta^y_1, \Delta^x_2, \Delta^y_2, \dots}$ that have a vanishing first-order 
term in \eq{equ:FirstOrderCondition} and thus for collective infinitesimal 
displacements $\Deltavec$ of all disks that escape from the packing. The script 
asserts that the configurations $\xvec+ 10^{-8}\Deltavec_a$ are without
overlaps and 
that all contacts persist at a precision $10^{-8}$ (contacts are lost at second 
order only). Furthermore, the script ensures that the uniform translations 
of all disks along the $x$- and $y$-axis are part of the solution 
space. Finally, the basis vectors $\Deltavec_a$ are stored in a 
human-readable output file, and optionally represented as in 
\fig{fig:Escape}. The input filename of the packing, and the output filename 
for the collective escape modes are specified in command-line arguments. 
Further optional arguments specify the filename for the plots of the escape 
modes, and the system length of the central simulation box (that is 
unnecessary for packings generated by the Python script 
\texttt{construct\_packing.py} in which case the system length is parsed from 
the packing file). The package's \texttt{README.md} 
file, as well as the \texttt{--help} command-line option and the 
docstrings of the script contain more detailed information.

\subsection{Go application \texttt{go-hard-disks}}

The Go application \texttt{go-hard-disks} relies on a cell-occupancy system for 
the efficient simulations of large-$N$ hard-disk systems using several variants 
of 
the ECMC algorithm. Straight, reflective, forward, and Newtonian ECMC are 
implemented. In its current form, it samples the maximum nearest-neighbor 
distance $d(t)$ (see \eq{equ:MaxNear}) after a given sampling time. All 
computations use a fixed number of mantissa bits (in base $2$) that may exceed 
the usual 24 or 53 bits for single- or double-precision floating-point values. 
(We use the \texttt{math/big} package of the Go standard library for the 
arbitrary-precision arithmetic.) The ECMC variant, its parameters (as for 
example
the sampling time or chain time), and further specifications 
(the number of mantissa bits, the cell specifications, the filename for the 
initial configuration, \etc) are again set using command-line arguments (see 
the 
\texttt{README.md} file of the package for  details on the installation process 
and the possible arguments).

\subsection{License, dependencies, software versions}

The \texttt{BigBoro} software package is published as an open-source project 
under the GNU GPLv3 license. It is available on GitHub as part of the 
\texttt{JeLLyFysh} organization.\footnote{The url of  repository is 
\url{https://github.com/jellyfysh/BigBoro}.} Users can clone or fork the 
repository to study the code, and to run the Python3 scripts 
\texttt{construct\_packing.py} and \texttt{collective\_escape\_modes.py}, and 
the Go application \texttt{go-hard-disks}. The Python3 scripts rely on NumPy as 
their only external dependency~\cite{numpy}. Optional plotting also requires
the Matplotlib library~\cite{matplotlib}. They are expected to work with any 
Python3 version and any NumPy version $\geq$ 1.20. Python scripts were tested 
with Python 3.9 and NumPy 1.21. The Go application only requires the Go 
standard library. It is expected to work with any Go version $\geq$ 1.13 (tested 
with $1.16$). Users can communicate with the authors (for suggestions or bug 
reports, \etc) through GitHub issues, and are encouraged to contribute to the 
project by pull requests.

\bibliographystyle{spphys}       

\bibliography{General.bib,Boroczky.bib} 

\end{document}